\documentclass[aps,twocolumn,prb,superscriptaddress,showpacs]{revtex4}%
\usepackage{graphicx}
\usepackage{amsfonts}
\usepackage{amsmath}
\usepackage{amssymb}
\usepackage{bm}
\usepackage{hyperref}%
\hypersetup{colorlinks,citecolor=blue,filecolor=blue,linkcolor=blue,urlcolor=blue}
\begin{document}
\title{Statistically induced topological phase transitions in one-dimensional superlattice anyon-Hubbard model}
\author{Zheng-Wei Zuo}
\author{Guo-Ling Li}
\author{Liben Li}
\affiliation{School of Physics and Engineering, and Henan Key Laboratory of Photoelectric Energy Storage Materials and Applications, Henan University of Science and Technology,
Luoyang 471003, China}
\date{\today }

\begin{abstract}
We theoretically investigate topological properties of the one-dimensional superlattice anyon-Hubbard model, which can be mapped to a superlattice bose-Hubbard model with an occupation-dependent phase factor by fractional Jordan-Wigner transformation. The topological anyon-Mott insulator is identified by topological invariant and
edge modes using exact diagonalization and density-matrix renormalization-group algorithm. When only the statistical angle is varied and
all other parameters are fixed, a statistically induced topological phase transition can be realized,
which provides new insights into the topological phase transitions. What's more, we give an
explanation of the statistically induced topological phase transition. The topological anyon-Mott phases can also appear in a variety of superlattice anyon-Hubbard models.
\end{abstract}


\maketitle

\section{Introduction}

Recently, one-dimensional (1D) topological quantum systems have attracted
increasing
attentions\cite{SchnyderAP08PRB,Kitaev09AIP,FidkowskiL10PRB,FidkowskiL11PRB,TurnerAM11PRB,ChenXie11PRB1,ChenXie11PRB2}%
. According to their symmetries, an exhaustive classification of all
topological phases of noninteracting fermions has been proposed in
Refs.\onlinecite{SchnyderAP08PRB,Kitaev09AIP}. The Su-Shrieffer-Heeger (SSH) model\cite{SuWP79PRL}, The Aubry-Andr\'{e}-Harper (AAH) model\cite{LangLJ12PRL,GaneshanS13PRL}, and the Kitaev chain\cite{Kitaev01} are three
prototypes of 1D noninteracting topological systems in this classification.
However, in the presence of strong interactions, the free classification in one-dimensional
is radically modified, which breaks down from $Z$ to $Z8$, i.e., there are
only eight distinct survived phases\cite{FidkowskiL10PRB}. So far, many different
analytical and numerical methods such as bosonization technique\cite{GogolinAO98Book,GiamarchiT04Book},
group cohomology\cite{ChenXie11PRB1}, exact diagonalization\cite{ZhangJM10EJP,RaventosD17JPB},
and the density matrix renormalization group (DMRG) algorithm have been used to study the 1D topological interacting systems,
including Luttinger liquid systems\cite{RuhmanJ15PRL,Keselman15PRB,KainarisN15PRB,MontorsiA17PRB,RuhmanJ17PRL},
interacting Floquet systems\cite{PotterAC16PRX,KeyserlingkCW16PRB,KeyserlingkCW16PRB2,DominicVE16PRB},
fermion-Hubbard systems\cite{XuZH13PRL,XuZH13PRB,GuoHM15PRB}, and bose-Hubbard systems\cite{ZhuSL13PRL,GrusdtF13PRL,DengXL14PRA,LiTH15PRB,MatsudaF14JPSJ,ZengTS16PRB}.

The classification of elementary particles as bosons and fermions is crucial
to the understanding of a variety of physical systems. However, in low
dimensions, particles with others kinds of quantum statistics, anyons, are
possible. Because of a wide range of unexpected properties, anyons play a more
and more important role in studies of topological order phases of
matter\cite{Stern10NT,SternA16ARCMP} and topological quantum
computation\cite{Nayak08RMP}. The 1D anyons systems have exotic and complicated
physics properties such as anyonic Bloch oscillations\cite{LonghiS12PRB}%
, asymmetric momentum distributions\cite{HaoYJ09PRA,TangGX15NJP}, and
statistically induced quantum phase transitions\cite{KeilmannT11NTC}.
Particularly interesting is the recent proposal of the 1D anyon-Hubbard
model\cite{KeilmannT11NTC,GreschnerS14PRL,WrightTM14PRL,Greschner15PRL,StraterC16PRL,Arcila-Forero16PRA,CardarelliL16PRA,TangGX15NJP,LangeF17PRL,LangeF17PRA,ZhangWZ17PRA}
in optical lattice. Various experimental realizations of anyon-Hubbard model
have been
proposed\cite{KeilmannT11NTC,GreschnerS14PRL,Greschner15PRL,StraterC16PRL},
from Raman-assisted tunneling scheme to lattice shaking-induced resonant
tunneling against potential offsets scheme. One of the great advantages of
these schemes is the experimental possibility to tune all parameters at will.
Theoretical and numerical studies show that the anyon-Hubbard model has a rich
physics. Besides the Mott insulator and superfluids, the pair superfluids,
dimer phases, and exotic partially paired phase are found\cite{Greschner15PRL}%
. In addition, the symmetry protected topological anyon-Haldane phase emerges
from the extended anyon-Hubbard model\cite{LangeF17PRL}. Now, the
anyon-Hubbard model receives a continuous and extensive interest.

As we know, the superlattice potential\cite{LangLJ12PRL,GaneshanS13PRL} and
electron interaction\cite{FidkowskiL10PRB,ZhuSL13PRL,GrusdtF13PRL} can induce
the topological phase transition. The interplay of the anyonic statistics,
superlattice potential, and interaction may exhibits a
rich physics of quantum matter. In this paper, we explore the nontrivial
topological properties of 1D anyon-Hubbard model with
superlattice potential by exact diagonalization\cite{ZhangJM10EJP,RaventosD17JPB}
and DMRG algorithm\cite{White92PRL,Schollwock05RMP,Schollwock11AP}. The paper is organized as
follows. Firstly, we study the topological properties of the Mott insulator
phase of superlattice anyon-Hubbard model at strong interaction strength, the
topological anyon-Mott insulator is demonstrated by topological invariant and
edge modes. Next, we consider the effect of statistical angle on the physical
properties of the superlattice anyon-Hubbard model. A statistically induced
topological phase transition occurs when the statistical angle increases and
other parameters are fixed. In addition, we show the topological phase
transition also appears in the off-diagonal superlattice anyon-Hubbard model.
Finally, a summary is given.

\section{Model and Result}

Here, we briefly introduce the 1D superlattice anyon-Hubbard model

\begin{equation}
H^{a}=-J\sum_{j}^{L-1}\left(  a_{j}^{\dagger}a_{j+1}+h.c.\right)  +\sum
_{j}^{L}\left[  \frac{U}{2}n_{j}\left(  n_{j}-1\right)  +V_{j}n_{j}\right]
\label{AHM}%
\end{equation}
where $J$ is the hopping strength, $U$ the on-site two-body interaction
energy. $V_{j}=V\cos\left(  2\pi\alpha j+\delta\right)  $ stands for a
periodic superlattice potential with $V$ being the modulation amplitude, and
$\delta$ being an arbitrary phase. We consider a commensurate
superlattice potential $V_{j}$ with modulation period $\alpha=p/q$ ($p,q$ are
integers) being a rational number. Here, $a_{j}^{\dagger}$ ($a_{j}$) is the
creation (annihilation) operator of anyon on site $j$, $n_{j}=a_{j}^{\dagger
}a_{j}$ is the number operator. These anyons satisfy the generalized
commutation relations
\begin{equation}
a_{j}a_{k}^{\dagger}-e^{-\theta sgn\left(  j-k\right)  }a_{k}^{\dagger}%
a_{j}=\delta_{kj},a_{j}a_{k}=e^{\theta sgn\left(  j-k\right)  }a_{k}a_{j}
\label{Anyon}%
\end{equation}
where $\theta$ denotes the statistical phase, and the sign function
$sgn\left(  j-k\right)  =\left(  j-k\right)  /\left\vert j=k\right\vert $ for
$j\neq k$, and $sgn\left(  j-k\right)  =0$ for $j=k$, respectively.

By fractional Jordan-Wigner transformation $a_{j}=b_{j}\exp\left(  i\theta
\sum_{i=1}^{j-1}n_{i}\right)  $, the superlattice anyon-Hubbard Hamiltonian
can be rewritten as a superlattice Bose-Hubbard model with an
occupation-dependent phase factor
\begin{align}
H^{b}= &  -J\sum_{j}^{L-1}\left(  b_{j}^{\dagger}b_{j+1}e^{i\theta n_{j}%
}+h.c.\right)  \nonumber\\
&  +\sum_{j}^{L}\left[  \frac{U}{2}n_{j}\left(  n_{j}-1\right)  +V_{j}%
n_{j}\right]  \label{BHM}%
\end{align}
where $b_{j}^{\dagger}$ is a boson creation operator. Due to the
occupation-dependent hopping, the reflection parity symmetry is broken. So, we
map the superlattice anyon-Hubbard model to an occupation-dependent hopping
superlattice boson-Hubbard model. The Hilbert space of anyons can be
constructed from that of bosons. For convenience, we take $J=1$ as the unit of
energy and choose $\alpha=1/3$.

As shown in Ref.\onlinecite{KeilmannT11NTC}, owing to the unitarity of anyon-boson
mapping, the two models are isospectral and they share the same energy gaps and
phase diagrams. For the statistical phase $\theta=0$, we can see that the
superlattice anyon-Hubbard model reduces to the superlattice boson-Hubbard
model\cite{ZhuSL13PRL}. The ground-state phase diagram of superlattice
boson-Hubbard model is
well-studied\cite{ZhuSL13PRL,DengXL14PRA,LiTH15PRB,MatsudaF14JPSJ,GuoHM15PRB}.
For the anyon case, we hope the interplay of anyonic statistics and
interaction lead to a rich physics of quantum phases. Without periodic
superlattice potential, the anyon-Hubbard model with repulsive interactions
has been observed that Mott insulator and superfluid phases characterized the
phase
diagram\cite{KeilmannT11NTC,GreschnerS14PRL,Greschner15PRL,StraterC16PRL,Arcila-Forero16PRA,CardarelliL16PRA,TangGX15NJP}%
. Here, we focus on the topological properties of the Mott insulator phase.

\subsection{Topological Anyon-Mott Insulator}

As we know, the superlattice anyon-Hubbard model is in superfluid phases for a
small $U$. As the interaction $U$ increases, the system transits into Mott
phase. Now, we investigate the topological properties of the Mott insulator
phase using DMRG algorithm and exact diagonalization.

Firstly, we tune the parameters of Hamiltonian to obtain the Mott state, which
can be characterized by the energy gap (particle-hole excitation)\cite{Arcila-Forero16PRA,RaventosD17JPB}, which is difference between quasiparticle and quasihole energy spectra. The quasiparticle and quasihole energy spectra can be defined as
\begin{align}
\mu_{p}(N)  &  =E_{N+1}-E_{N}\label{quasiParticle}\\
\mu_{h}(N) &  =E_{N}-E_{N-1} \label{quasihole}%
\end{align}
where $E_{N}$ is the ground-state energy of the system with $N$ anyons. So,
the energy gap $\Delta\mu$ can be obtained via
\begin{equation}
\Delta\mu=\mu_{p}(N)-\mu_{h}(N)=E_{N+1}+E_{N-1}-2E_{N} \label{ChemPotential}%
\end{equation}

\begin{figure}[ptbh]
\centering{ \includegraphics[scale=0.44]{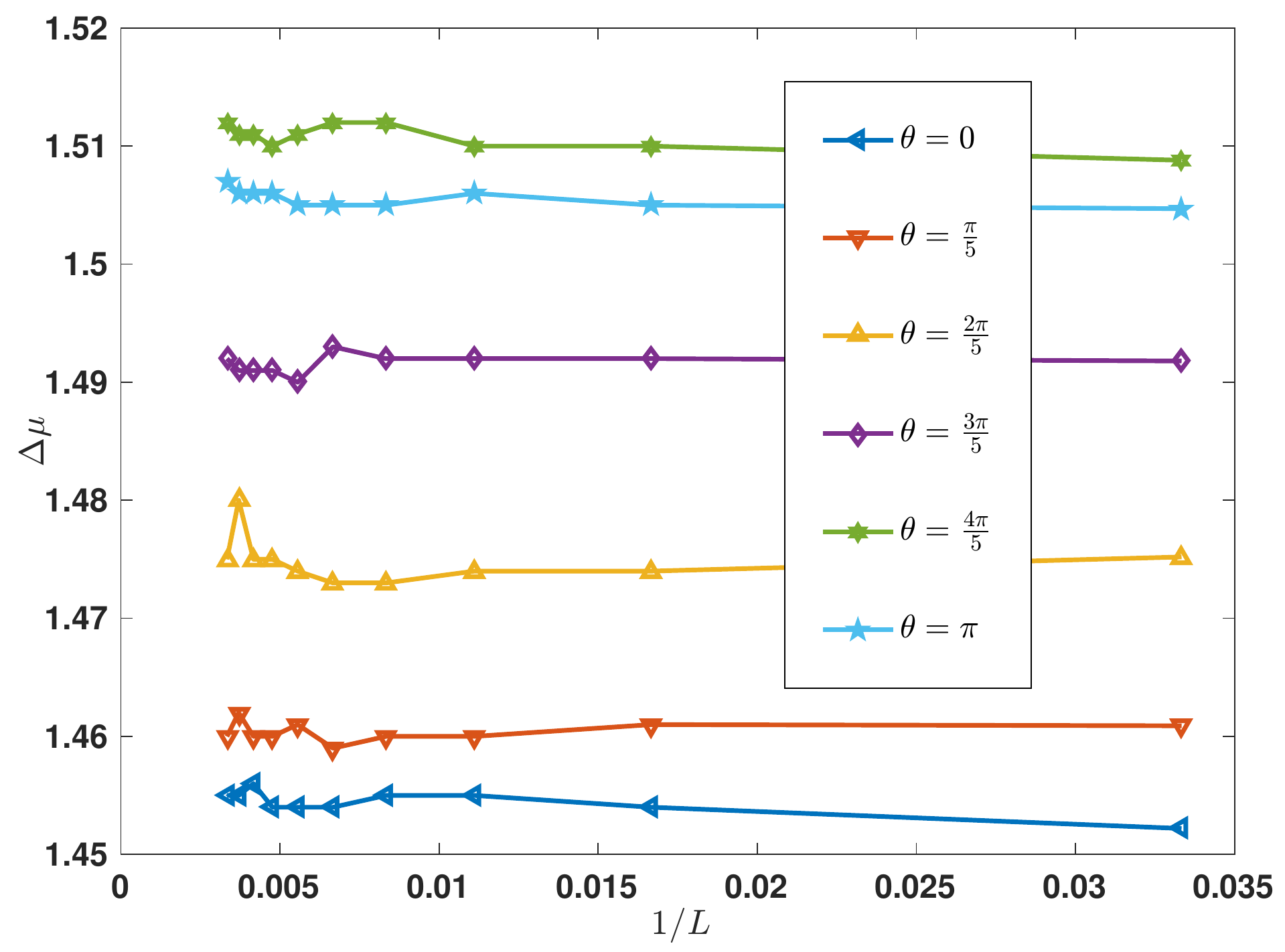}}%
\caption{(Color online) System size dependence of the gap $\Delta\mu$ of
anyon in anyon-Hubbard superlattice with statistical angle $\theta
=0,\pi/5,2\pi/5,3\pi/5,4\pi/5,\pi$, $U=10$, $V=10$,$\delta=2\pi/3$, $\alpha
=\nu=1/3$.}%
\label{FigChemPotential10}%
\end{figure}

At the thermodynamic limit,  the gap $\Delta\mu$ is finite and zero in Mott  and superfluid
phases, respectively. We start considering the superlattice anyon-Hubbard model  with the
number of anyons $N$ and the lattice sites $L$. The filling factor is defined as $\nu=N/L=m\alpha$ with $m$ being an
integer. Here, we set $\alpha=\nu=1/3$, $m=1$, $U=10$, $V=10$, $\delta=2\pi/3$. Using the DMRG
algorithm based on the ITensor library\cite{ITensor}, we numerically get the
system size dependence of the gap $\Delta\mu$ of anyon in anyon-Hubbard superlattice with statistical
angle $\theta=0,\pi/5,2\pi/5,3\pi/5,4\pi/5,\pi$ with period boundary condition (PBC), which is
shown in Fig. \ref{FigChemPotential10}. The maximal boson number per site in DMRG is constrained to be four.

From Fig.\ref{FigChemPotential10}, we can see that the gap $\Delta\mu$ remain finite
values for $\theta=0,\pi/5,2\pi/5,3\pi/5,4\pi/5,\pi$ at the thermodynamic
limit. These results suggest that the ground states correspond to Mott
insulator phases for any statistical angle $\theta$. In the following
sections, we mainly focus on the topological feature of superlattice
anyon-Hubbard model for $0\leqslant\theta\leqslant\pi$.

The topological property of these Mott states can be demonstrated by the Chern
number for many-body interacting system\cite{NiuQ85PRB}. First, let us introduce the twisted boundary condition
$\left\vert \psi\left(  j+L,\delta,\phi\right)  \right\rangle =e^{i\phi
}\left\vert \psi\left(  j,\delta,\phi\right)  \right\rangle $ where $j$
denotes an arbitrary site, $\alpha$ is the twist angle and takes values from
$0$ to $2\pi$. Then, the Chern number $C$ as a topological invariant can be
calculated by the following formula
\begin{equation}
C=\frac{1}{2\pi}\int_{0}^{2\pi}\int_{0}^{2\pi}F\left(  \delta,\phi\right)
d\delta d\phi
\end{equation}
where $F\left(  \delta,\phi\right)  =\operatorname{Im}\left(  \left\langle
\frac{\partial\psi}{\partial\phi}\left.  {}\right\vert \frac{\partial\psi
}{\partial\delta}\right\rangle -\left\langle \frac{\partial\psi}%
{\partial\delta}\left.  {}\right\vert \frac{\partial\psi}{\partial\phi
}\right\rangle \right)  $ is the Berry curvature and $\psi$ many-body
ground-state wave function of the system. The Chern number can be interpreted
as numbers of anyon charge pumped in one cycle of $\delta$ for the superlattice
anyon-Hubbard model. We numerically calculate the Chern
numbers of the ground states. By using the exact diagonalization and the
discrete set of values of $\left(  \delta,\phi\right)  $, we calculate the
Chern number $C$\cite{FukuiT05JPSJ} with system size L=15. We find that the Chern number $C=1$. These
calculations demonstrate that these Mott states are topological. The
topological properties of the Mott states with other statistical angles can be
analyzed similarly. We refer to these topological Mott states as
\textit{topological anyon-Mott insulator}. For the statistical angle
$\theta=0$, the results are consistent with those calculations for
superlattice Bose-Hubbard model and the topological Bose-Mott insulator is
obtained\cite{ZhuSL13PRL}.

\begin{figure}[ptbh]
\centering{ \includegraphics[scale=0.4]{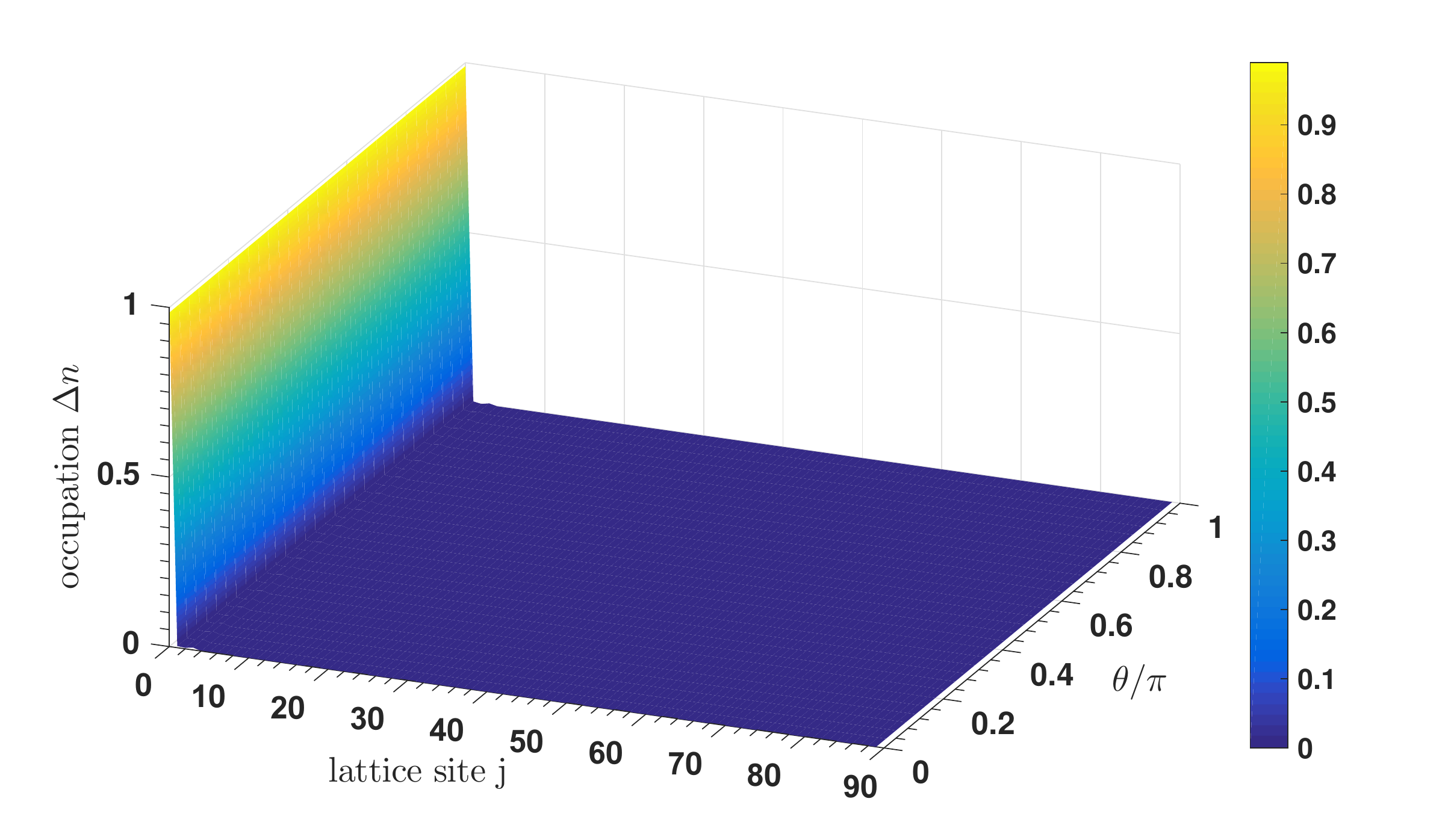}}%
\caption{(Color online) The density distribution of the quasiparticle along
the lattice sites at $N=30,L=90$. The other parameters are the same as those
for Fig.\ref{FigChemPotential10}.}%
\label{FigDistribution10}%
\end{figure}

According to the bulk-edge correspondence, there are edge modes (localized
modes) in topological interacting systems characterized by the quasiparticle
density
distribution\cite{ZhuSL13PRL,GrusdtF13PRL,DengXL14PRA,LiTH15PRB,MatsudaF14JPSJ,GuoHM15PRB}%
. The density distribution of the quasiparticle can be defined as%
\begin{equation}
\Delta n_{j}=\left\langle \psi_{n+1}^{g}\left\vert n_{j}\right\vert \psi
_{n+1}^{g}\right\rangle -\left\langle \psi_{n}^{g}\left\vert n_{j}\right\vert
\psi_{n}^{g}\right\rangle \label{distributionQP}%
\end{equation}
where $\psi_{n}^{g}$ denotes the ground state wave function of the system with
$n$ bosonic atoms for open boundary condition (OBC), the quasihole case has similar
formula. The results with different statistical angles $\theta$ for $N=30,L=90$ using DMRG algorithm are
shown in Fig. \ref{FigDistribution10}. Such edges modes which mainly distribute near one end site of the chain can be seen in Fig. \ref{FigDistribution10}, which are consistent with the topological invariant Chern numbers.

\subsection{Statistically Induced Topological Phase Transition}

\begin{figure}[htbp]
\centering{ \includegraphics[scale=0.45]{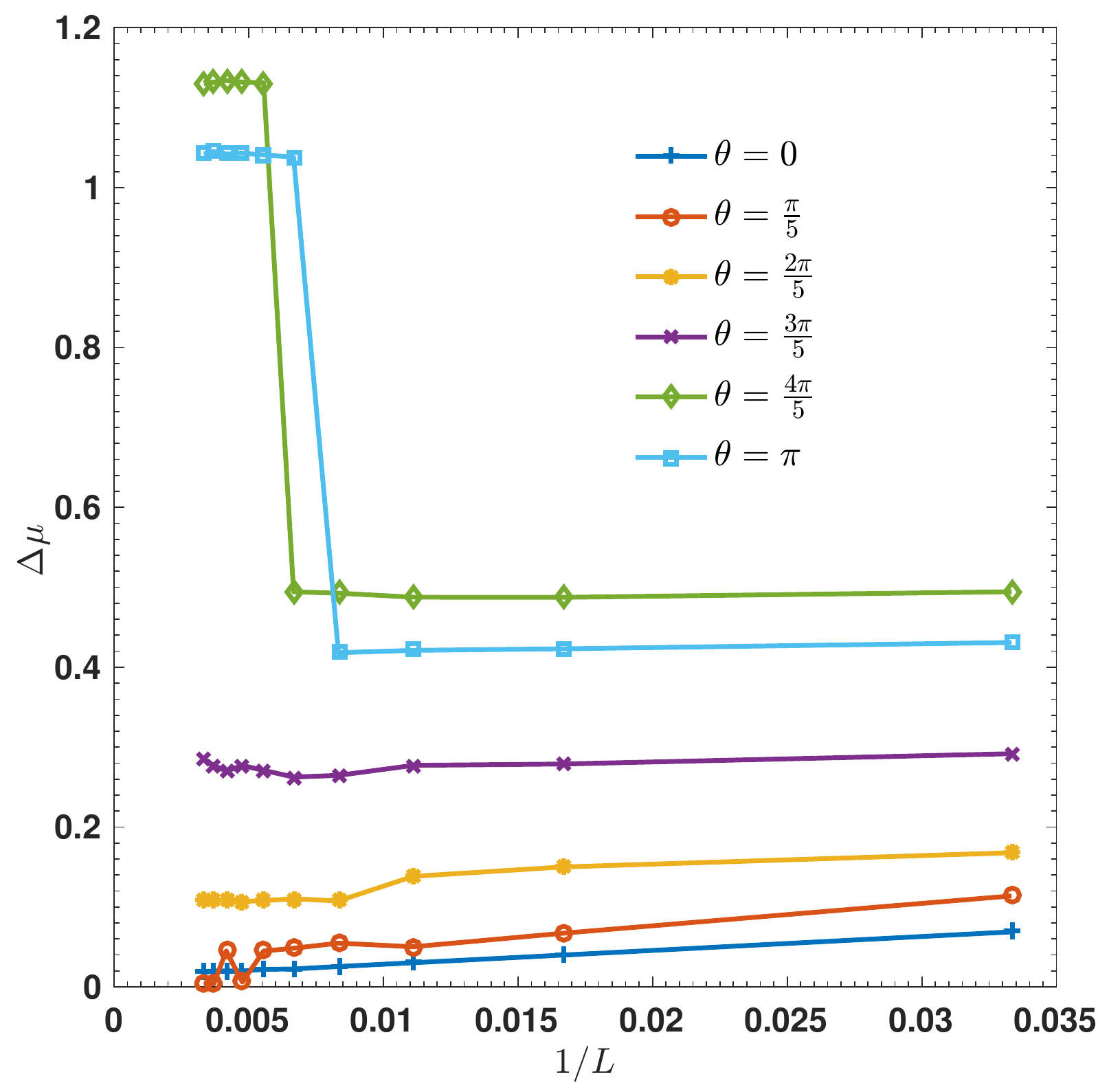}}\caption{(Color
online) System size dependence of the gap $\Delta\mu$ of anyon in
anyon-Hubbard superlattice with statistical angle $\theta=0,\pi/5,2\pi
/5,3\pi/5,4\pi/5, \pi$, and $U=2$, $V=1$, $\delta=2\pi/3$,
$\alpha=\nu=1/3$.}%
\label{FigChemPotential1}%
\end{figure}

Now, we consider the effect of statistical angle on the physics properties of
the superlattice anyon-Hubbard model. Here, we take the parameters $U=2$, $V=1$, $\delta=2\pi/3$, $\alpha=\nu=1/3$ as an example. To
begin with, the evolution of the gap $\Delta\mu$ versus the inverse of the
lattice length appear in Fig. \ref{FigChemPotential1}. At the thermodynamic limit, the
value of the gap $\Delta\mu$ is zero and the ground state is superfluid at
$\theta=0$. However, at $\theta=4\pi/5,\pi$, the gap $\Delta\mu$ remain finite
value and the ground states are in Mott state. This figure suggests that the
system transits from superfluid into Mott phase when the statistical angle
increases, so the phase transition takes place. Next, we demonstrate that these Mott states are topological.

\begin{figure}[htbp]
\centering{ \includegraphics[scale=0.45]{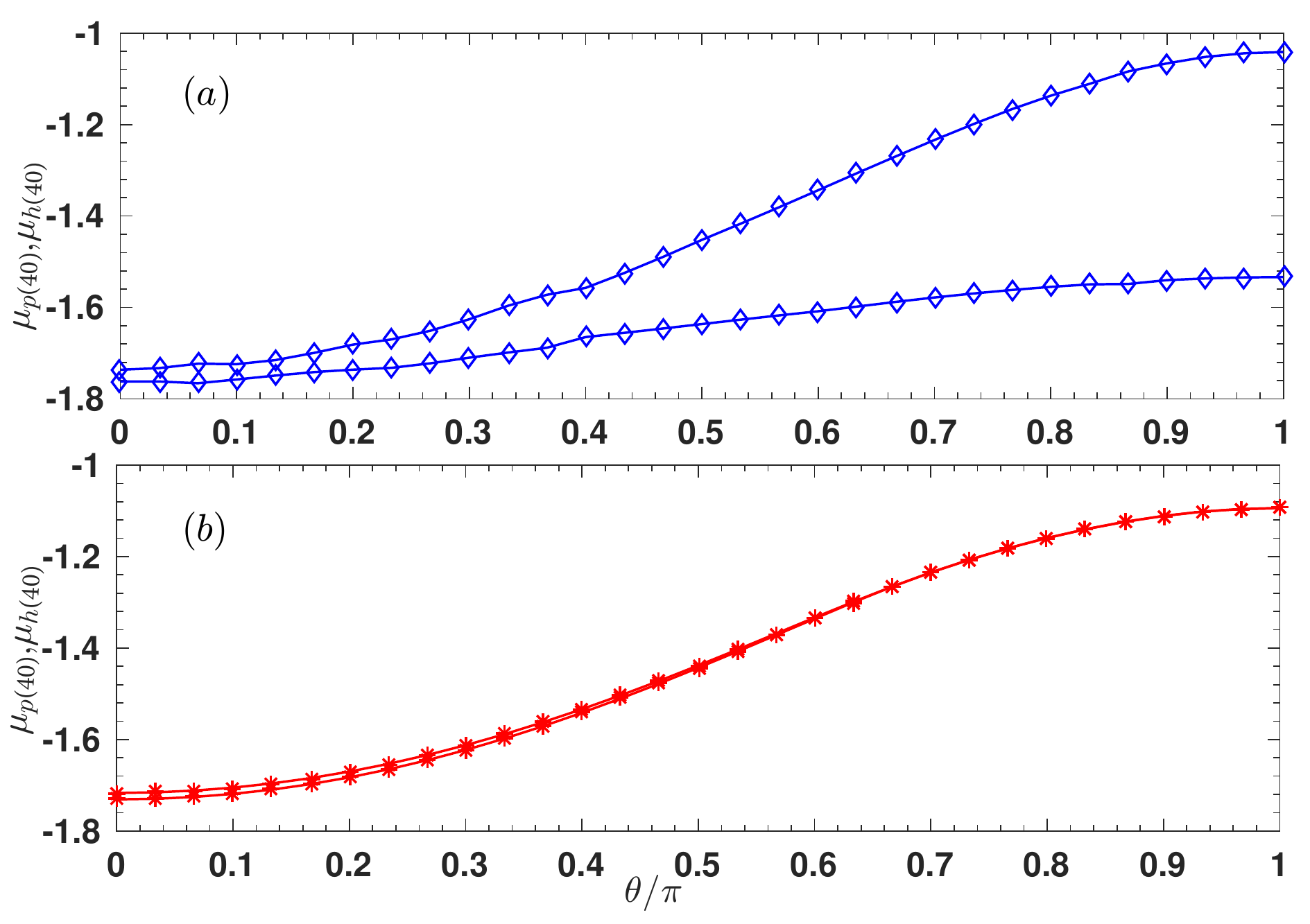}}\caption{(Color
online)The quasiparticle and quasihole energy spectra versus the statistical
angle $\theta$ for anyon-Hubbard superlattice with $L=120$, $U=2$,
$V=1$, $\delta=2\pi/3$, $\alpha=\nu=1/3$; $(a)$ Period boundary condition
$(b)$ Open boundary condition.}%
\label{FigChemicalGap}%
\end{figure}

When the system has a nontrivial topological property, there are states in the
gap of the quasiparticle energy spectrum as the boundary condition changes
from PBC to OBC\cite{GuoHM11PRB}. Now, the calculations are taken that quasiparticle
and quasihole energy spectra versus the statistical angle $\theta$ with
$L=120$, $U=2$, $V=1$, $\delta=2\pi/3$, $\alpha=\nu=1/3$ for (a) PBC
and (b) OBC, which are shown in Fig. \ref{FigChemicalGap}. This figure shows that
the gap $\Delta\mu$ becomes more and more larger for PBC when the  statistical angle increases.
For large $\theta$, the system transits into the Mott phase ($\Delta\mu$=0.492 for $\theta=\pi$), as
shown in Fig. \ref{FigChemPotential1}. The energies of quasiparticles added or removed (for OBC) appear in the gap (for
PBC) when the statistical angle exceeds a critical value. The two in-gap
modes have exactly the same values for larger statistical angle. For small $\theta$, the system
enters into the superfluid phase, as shown in Fig. \ref{FigChemPotential1}.
these two in-gap modes disappear and evolve into the bulk ones. The differences of quasiparticle
and quasihole energy spectra are very small for small $\theta$. For example,  the difference
of quasiparticle spectrum for PBC and OBC is $0.0193$ and the difference
of quasihole spectrum is $0.0308$.  These values should have a tendency to zero at the
thermodynamic limit because of the superfluid regime. These results show that
topological phase transition arises when the statistical angle is large. When the
statistical angle increases, the topological quantum transition occurs.
As we know, one of the weaknesses of the DMRG is that it works poorly with PBC. This stems from the fact that conventional DMRG optimizes over open-boundary matrix product state wave-functions. For the larger size, the accuracy of data maybe drastically lower under PBC. This may be why the data $\theta=4\pi/5$ and $\pi$ show weird behavior in Fig. \ref{FigChemPotential1}. So, the critical statistical angle of this topological quantum transition should be estimated by the data of DMRG with OBC and Chern number.

\begin{figure}[ptbh]
\centering{ \includegraphics[scale=0.43]{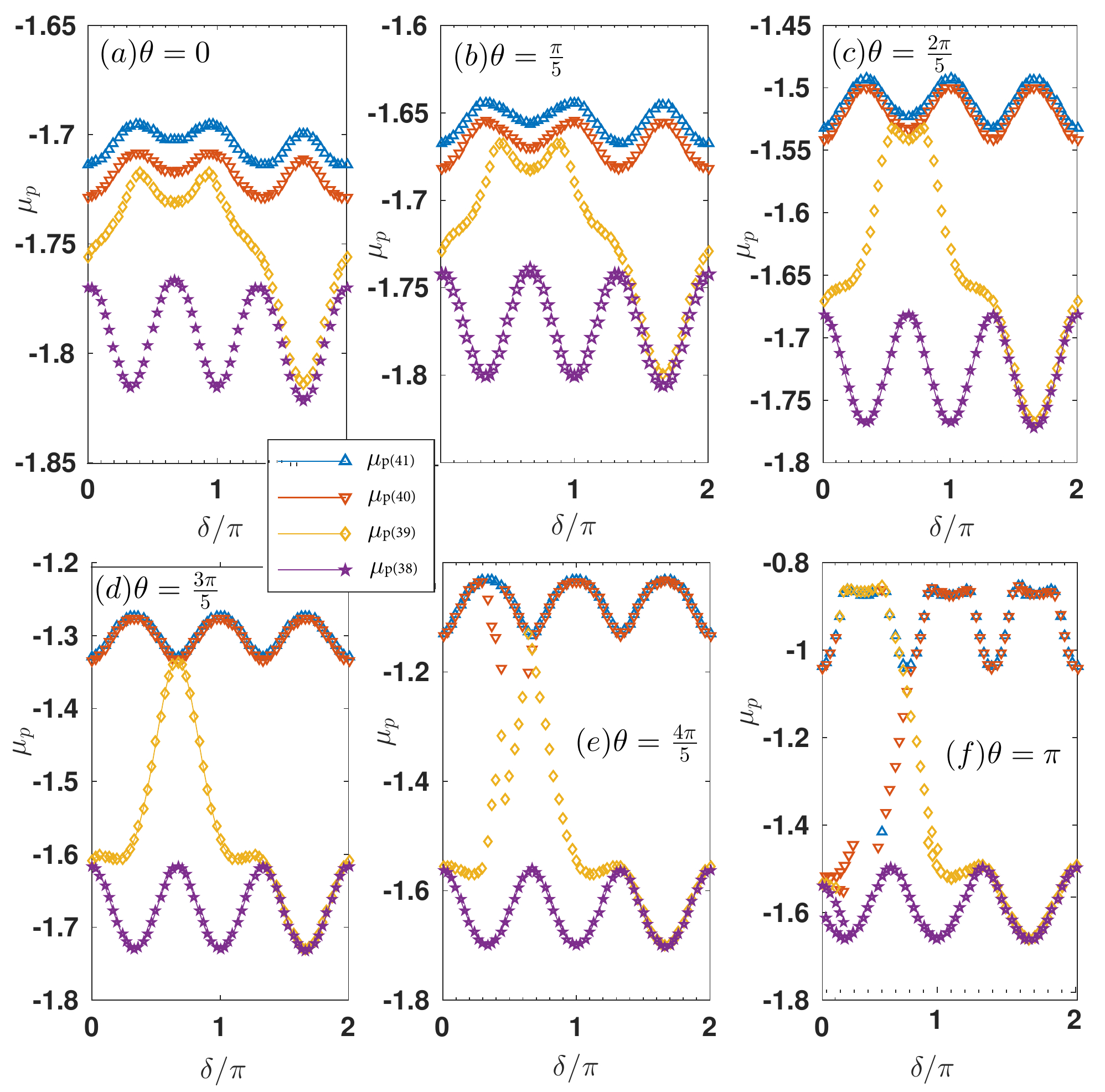}}\caption{(Color
online)The quasiparticle energy spectrum $ \mu_{p}(N)$ ($N=$41, 40, 39, and 38) with respect to
phase parameter $\delta$ for anyon-Hubbard superlattice with $L=120$, $U=2$, $V=1$,  
and  $\theta=0, \pi/5, 2\pi/5, 3\pi/5, 4\pi/5, \pi$ under OBC.}
\label{FigDeta}%
\end{figure}

To illustrate the topological quantum transition clearly,  we calculate quasiparticle energy spectrum with respect to the
phase parameter $\delta$ around the filling factor $\nu=1/3$  by way of DMRG algorithm under OBC. In Fig. \ref{FigDeta}, we
show the quasiparticle energy spectrum $\mu_{p}(N)$ ($N=$41, 40, 39, and 38)  as a function of the phase parameter $\delta$ with
$L=120$, $U=2$, $V=1$, and  $\theta=0,\pi/5,2\pi/5,3\pi/5,4\pi/5, \pi$.  As illustrated in the figures,  there is a quasiparticle finite
gap for every  $\theta$ when deviating from filling factor $\nu=1/3$ ($N=$41, 38). For the $\theta=3\pi/5,4\pi/5$, and $\pi$,  there
exist two branches of quasiparticle modes which cross each other and connect the lower and upper
bands of quasiparticle energy spectra as the  $\delta$  varies from $0$ to $2\pi$. However, for  $\theta=0,\pi/5$, and $2\pi/5$, the  two
quasiparticle modes do not connect the lower and upper bands.  From the Fig. \ref{FigDeta}, it is clear that with the increasing statistical angle, a topological quantum transition takes place and the critical statistical angle is around the $3\pi/5$.

\begin{figure}[ptbh]
\centering{ \includegraphics[scale=0.36]{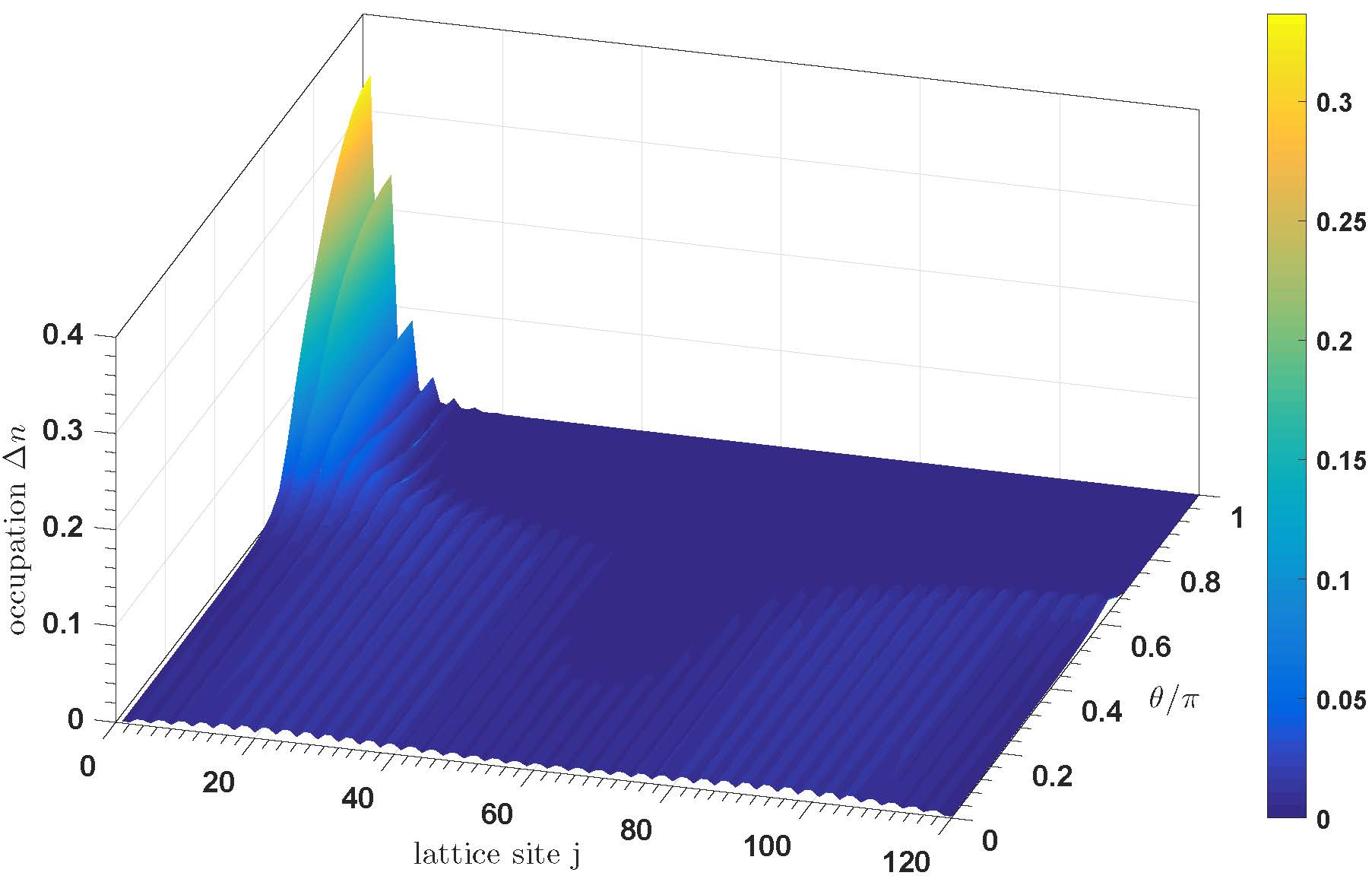}}\caption{(Color
online) The distribution of the quasiparticle along the lattice sites at
$L=120, 0\leqslant\theta\leqslant \pi$. The other parameters are the same as those for Fig.\ref{FigChemPotential1}.}%
\label{FigDistribution1}%
\end{figure}

Next, we verify that these crossing in-gap modes correspond to the end states and
calculate the quasiparticle density distribution of these modes, and then
compute the system's topological invariants. Here, we set the parameters $L=120$, $N=40$, $U=2$, $V=1$, and $\delta=2\pi/3$. According to Eq. (\ref{distributionQP}), the calculations are straightforward and the result (only quasiparticle
case) is shown in Fig. \ref{FigDistribution1}. As the statistical angle increases, the
density distribution begins to evolve from the bulk to the ends. For
$\theta>9\pi/10$, we can see that more than $30\%$ of the quasiparticle modes are
localized at the end sites, which are hallmark of topologically nontrivial
phase. The larger statistical angle is, the more probability of quasiparticle
are localized at the ends. From Fig. \ref{FigDistribution1}, we can see that the critical
statistical angle $\theta_{c}$ is about $0.65\pi$. The numerical results show that there 
are great probability distribution localized at the end sites when we adjust the phase 
parameter $\delta$ and/or periodic superlattice potential $V$ in topological anyon-Mott phase. Varying $\delta$ continuously 
from $0$ to $2\pi$, one can see that the end states change from one side to the other.

In the following, we calculate the topological invariants for the system with
$\theta=4\pi/5$ in the Mott regime. The topological property of the bulk Mott
state with concrete $\delta=2\pi/3$ can be characterized by the Berry phase
using the twisted boundary condition defined above. The Berry phase can be
defined as%
\begin{equation}
\gamma=
{\displaystyle\oint}
i\left\langle \psi\left(  \phi\right)  \left\vert \frac{d}{d\phi}\right\vert
\psi\left(  \phi\right)  \right\rangle
\end{equation}

\begin{figure}[ptbh]
\centering{ \includegraphics[scale=0.44]{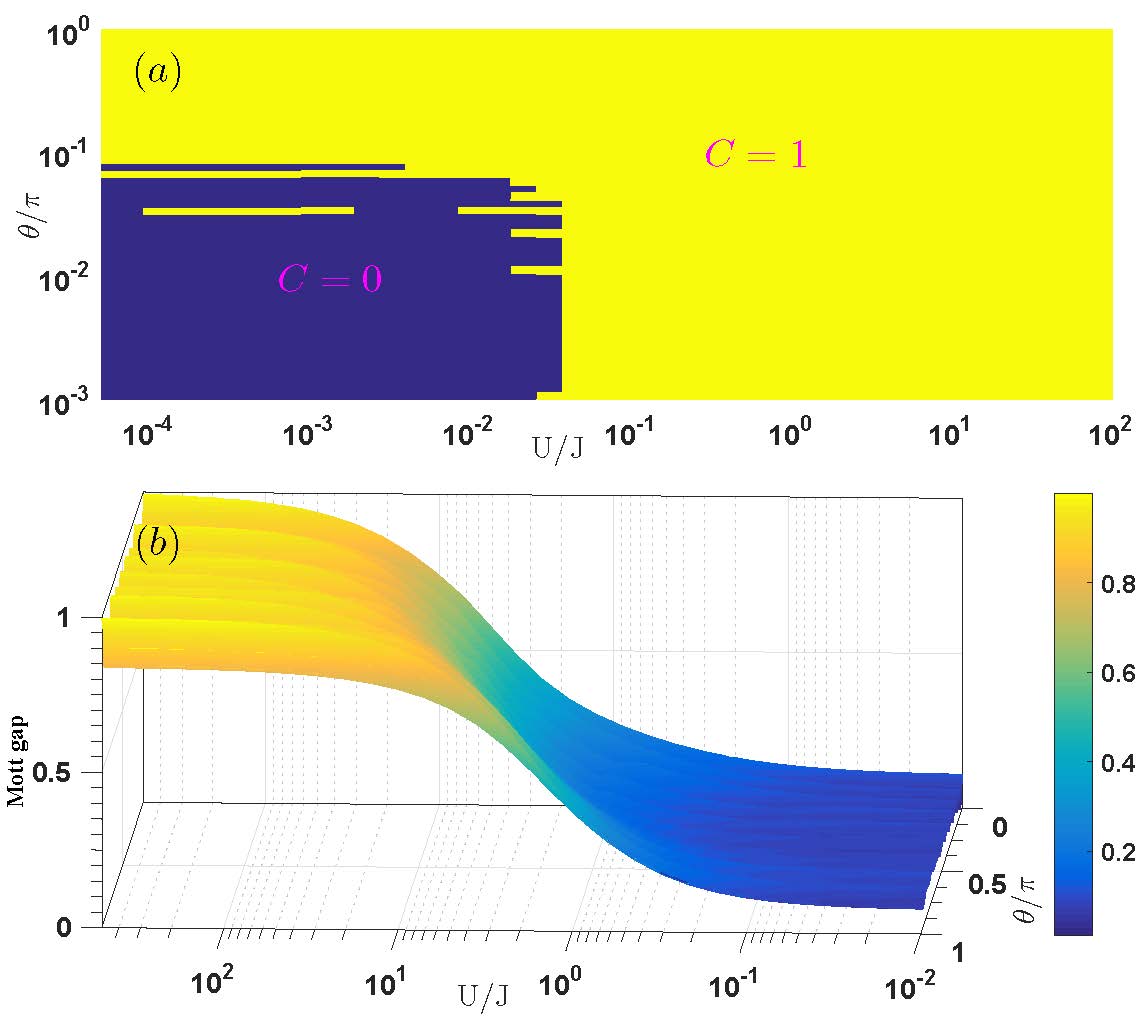}}\caption{(Color
online) $(a)$ The phase diagram for the system with $L=15$, $N=5$, and $V=1$; $(b)$ The Mott gap of the system with $L=18$, $N=6$, $V=1$ and $\delta=2\pi/3$.}
\label{PhaseDiag}%
\end{figure}

We find that the Berry phase $\gamma=\pi$, which shows the topological
property of the Mott state is nontrivial. In some region of the superfluid
(for small $\theta$) phase, Berry phase is still approximately $\pi$, because
of the finite-size gap. In addition, we can calculate the Chern number of the
system with $\theta=4\pi/5$ within $\left(  \delta,\phi\right)  $-space and
find the Chern number is quantized as $C=1$. This topological Mott state is
also the topological anyon-Mott insulator state. With fixed $J$, $U$, and $V$,
this phase transition is only driven by the statistical angle $\theta$. Taking
into account the above discussion and these numerical results, we refer this
phase transition as \textit{statistically induced topological phase
transition}. In principle, we can get the phase diagram by calculating the Chern number
or Berry phase. Here, we give the phase diagram for the system with $L=15$, $N=5$, and
$V=1$ using the Chern number and the Mott gap between ground and first excited
state of the system with $L=18$, $N=6$, $V=1$, and $\delta=2\pi/3$, which are
shown in Fig.\ref{PhaseDiag} $(a)$ and $(b)$. From Fig.\ref{PhaseDiag} $(a)$, we
can see that in some superfluid region, Chern numbers remain the value of 1 due
to the finite-size gap. Although, we cannot obtain the precise phase diagram by the 
Chern number,  the Mott gap can approximately reflect the evolution behaviors of 
the system by exact diagonalization. For every statistical angle, the Mott gap tends to a saturation 
value and zero at larger and smaller $U$, respectively, which correspond to the Mott and 
superfluid regions as shown in Fig.\ref{PhaseDiag} $(b)$ . The larger the statistical angle becomes, the
earlier the system enters into the topological anyon-Mott phase. With the statistical angle
increasing, the critical value of the superfluid to topological
anyon-Mott phase transition becomes smaller and smaller. Thus, for a fixed $U$
(for example $U=2$), the system can transfer from the superfluid phase into the topological
anyon-Mott phase as the statistical angle increases. In a word, as the value of
statistical angle is varied, the statistically induced topological phase transition emerges.

The statistically induced topological phase transition can be understood by the
following argument:  Using bosonization technique, Greschner and 
coauthors have shown that the anyonic exchange statistics has qualitatively the same effect as repulsive interaction in anyon-Hubbard with weakly interaction regime\cite{GreschnerS14PRL,Greschner15PRL}.  The effect of anyonic exchange statistics can be described by a scattering length that describes repulsive interactions.  With increasing anyonic angle, the effect scattering length increases which shows that the repulsive interaction becomes bigger. So anyonic exchange statistics and
repulsive interaction may have the same effect on phase diagram. The Mott lobes grows as the statistical angle increases, as shown by previous
results\cite{KeilmannT11NTC,GreschnerS14PRL,Greschner15PRL,Arcila-Forero16PRA,ZhangWZ17PRA}. Thus, we can take  the statistical
angle as an effective repulsion interaction parameter.  What's more, the
occupation-dependent phase factor in Eq. (\ref{BHM}) becomes more and more
important with increasing $\theta$. Because of the incoherent superpositions,
nearest neighbor tunneling processes would cancel out in the kinetic
Hamiltonian and contribute different values. The increase of statistical angle
enhances the destructive interference effect\cite{KeilmannT11NTC}. At the same
time, anyons with statistics angle $\theta=\pi$ are pseudofermions, because
they are fermions off-site, while being bosons on-site.  For large statistical angle $\theta\approx\pi$, the generalized off-site
commutation relations of anyons tend to generate a Pauli-exclusion principle.
As a result, the particles are more localized and an insulating phase would
appear. On the other hand, the bose interaction can induce the topological phase
transition in superlattice Bose-Hubbard model. So, the anyonic exchange statistics
can also drive the system into topological anyon-Mott insulator states as we shown above.

\subsection{Off-diagonal Anyon-Hubbard Superlattice}

\begin{figure}[tbh]
\centering{ \includegraphics[scale=0.4]{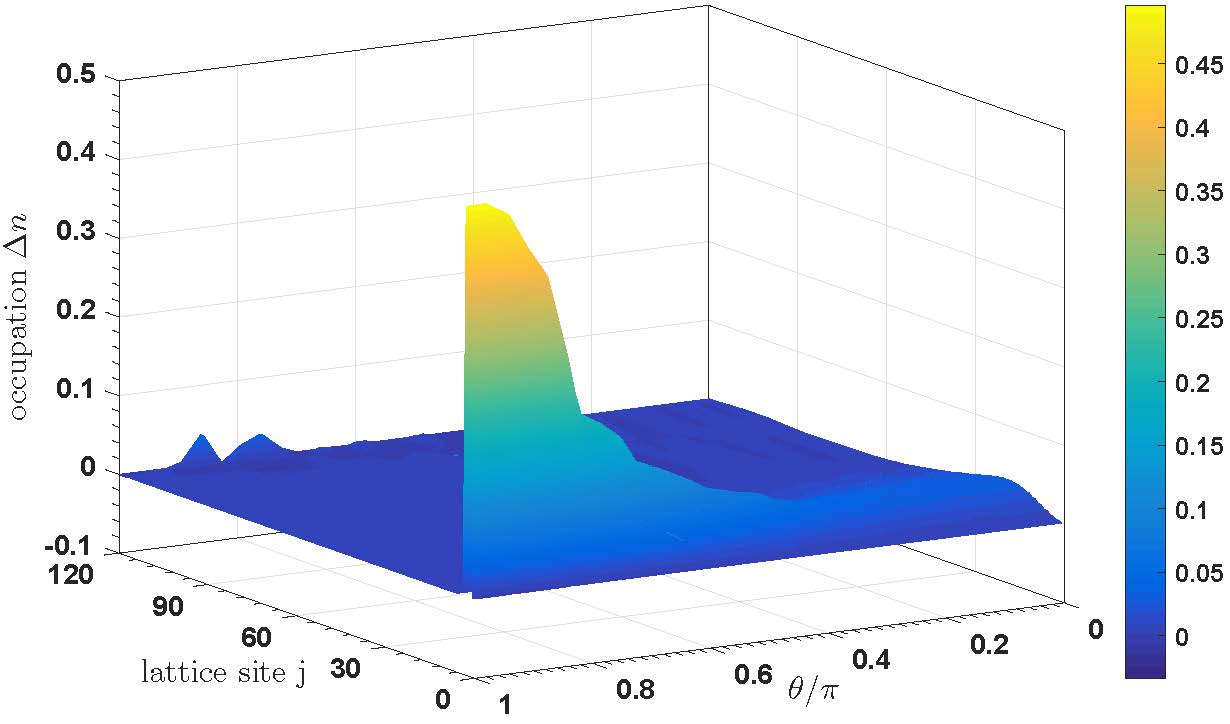}}%
\caption{(Color online) The distribution of the quasiparticle along the
lattice sites at $L=120$. The other parameters are $\alpha=\nu=1/2$, $U=1$,
$V=0.9$, $\delta=0$.}
\label{OffAAHDistribution}%
\end{figure}

As shown above, anyonic statistics and interaction result in a rather rich
physics in diagonal superlattice anyon-Hubbard model. In the following, we discuss
other types of superlattice anyon-Hubbard model. Here we consider 1D
off-diagonal superlattice anyon-Hubbard model, which can be defined as
\begin{equation}
H^{a}=-\sum_{j}^{L-1}\left(  J_{j}a_{j}^{\dagger}a_{j+1}+h.c.\right)
+\sum_{j}^{L}\frac{U}{2}n_{j}\left(  n_{j}-1\right)  \label{OffDiagnol}%
\end{equation}
where $J_{j}=\left[ 1+V\cos\left(  2\pi\alpha j+\delta\right) \right]$. According to the
above methods, we can also show that this off-diagonal superlattice
anyon-Hubbard model exhibits the topological anyon-Mott insulator phase and
the statistical angle can induce the topological phase transition. Without loss of generality, hereafter we
take the parameters $\alpha=1/2$, $\nu=1/2$, $U=1$, $V=0.9$, $\delta=0$, $L=120$
as an example. As the anyonic statistical angle increases, the edge modes begin to appear
and topological phase transition occurs, as shown by Fig. \ref{OffAAHDistribution} . What's more,  we numerically calculate the Berry
phase  and find $\gamma=\pi$ in topological anyon-Mott states. In a word, the statistically induced topological phase
transition exhibits in diagonal and off-diagonal superlattice anyon-Hubbard models.

\section{Conclusion}

In summary, we have researched the nontrivial topological properties of the
one-dimensional superlattice anyon-Hubbard model. For the strong interaction strength, the topological anyon-Mott
insulator is identified by the edge states and topological invariant. The anyonic exchange statistics affects significantly 
the ground-state properties of the system. When we take statistical angle as a free controllable parameter, the 
statistically induced topological phase transition appears, which
provides new insights on the topological phase transitions. Furthermore, we provide an explanation
of the statistically induced topological phase transition. In addition, our work
may open many directions for the superlattice anyon-Hubbard model. The superlattice anyon-Hubbard model can be generalized to a
variety of versions such as incommensurate superlattice potentials, non-Abelian, and non-Hermitian
cases. There would be a rich intriguing physics in these models.

\emph{Acknowledgements.}--This work was supported by the National Natural
Science Foundation of China under grant numbers 11604081, 11447008 (Z.Z.W.), U1404212 (G.L.L).

\end{document}